\documentclass[reprint,superscriptaddress,amsmath,amssymb,aps,pra]{revtex4-2}

\usepackage{graphicx}                                                                                                                                                                                                                                                                                             
\usepackage{dcolumn}
\usepackage{bm}

\begin{document}

\preprint{AIP/123-QED}

\title{Localization versus incommemsurability for finite boson system in one-dimensional disordered lattice}
\author{Barnali Chakrabarti}
\email{barnali.physics@presiuniv.ac.in}
\affiliation{Department of Physics, Presidency University, 86/1   College Street, Kolkata 700073, India.}
\affiliation{Instituto de Física, Universidade de São Paulo, CEP 05508-090, São Paulo-SP, Brazil.}

\author{Arnaldo Gammal}
\affiliation{Instituto de Física, Universidade de São Paulo, CEP 05508-090, São Paulo-SP, Brazil.}

\date{\today} 

\begin{abstract}
We explore the effect of disorder on a few-boson system in a finite one-dimensional quasiperiodic potential covering the full interaction ranging from uncorrelated to strongly correlated particles. We apply numerically exact multiconfigurational time-dependent Hartree for bosons to obtain the few-body emergent states in a finite lattice for both commensurate and incommensurate filling factors. The detailed characterization is done by the measures of one- and two-body correlations, fragmentation, order parameter. For commensurate filling, we trace the conventional fingerprints of disorder induced localization in the weakly interacting limit, however we observe robustness of fragmented and strongly correlated Mott in the disordered lattice. For filling factor smaller than one, we observe existing delocalization fraction of particles interplay in a complex way. For strongly interacting limit, the introduced disorder drags the fragmented superfluid of primary lattice to Mott localization. For filling factor larger than one in the primary lattice, the extra delocalization always resides on commensurate background of Mott-insulator. We observe beyond Bose-Hubbard physics in the fermionization limit when the pairing bosons fragment into two orbitals- Mott dimerization happens. The introduced disorder first relocates the dimers, then strong disorder starts to interfere with the background Mott correlation. These findings unlock a rich landscape of unexplored localization process in the quasiperiodic potentials and pave the way for engineering exotic quantum many-body states with ultracold atoms.
\end{abstract}


\maketitle

\section{Introduction} \label{sec:intro}
The flexible controlling in optical lattice enables the investigation of the most remarkable paradigmatic phase transition from coherent superfluid (SF) to localized Mott-insulator (MI) phase~\cite{Fisher:1989,nature.415,nature.419}. The MI characterizes exotic quantum phenomenon like fractional quantum hall effect~\cite{Burkov}, metal-insulator phase transition in the strongly correlated quantum matter~\cite{Theodor}, topological phase transition~\cite{Sudeshna}. It further includes many-body localization~\cite{Bloch:2019,Vadim}, Bose glass (BG) phases~\cite{Giamarchi:1988,Fisher:1989,Fallani:2005,Palencia:2019,Pasienski:2010}, fractal Mott lobes~\cite{Palencia:2019,Palencia:2020,yao2024}, Anderson localization~\cite{Lugan:2007,Lugan:2011,Billy:2008,Paliencia:2007,Lugan1:2007} in disordered lattice. 
Experimentally, disorder can be introduced by a bichromatic potential~\cite{Fallani:2005,Fallani:2007}. A quasiperiodic potential is
formed by a main optical lattice of high intensity superposed by a second weaker one with slightly
different wave length, the resulting lattice displays local inhomogeneities. The phase diagrams of quasiperiodic systems with two periodic lattices of incommensurate period have been theoretically studied both in one and two dimensions~\cite{Zhu,Roux,Yao:2021,Palencia:2020}. Controlled quasiperiodic potentials are also realized in the experiments with ultracold atoms~\cite{Roati:2008, Palencia:2010, Modugno:2010, Chiara,Chiara1}.  

The one-dimensional quasiperiodic lattice is the most versatile platform to understand localization versus delocalization. For the commensurate filling, interacting bosons in the regular periodic lattice exhibits incompressible Mott insulating phase, whereas for the incommensurate filling an extended superfluid with global correlation is expected. When bosons in the periodic lattice shows a transition from MI to SF phase, in the disordered lattice, the MI to SF transition is intercepted by a compressible but non-coherent phase-Bose Glass phase. In recent studies, the interplay between commensuration and localization has been studied both theoretically and experimentally~\cite{yao2024, BarnaliPRB}. It has been illustrated that when the number of particles commensurate with the number of lattice sites, a localized Mott phase in the periodic lattice turns to delocalized superfluid in the quasiperiodic lattice~\cite{yao2024}. It is claimed that the anomalous delocalization is introduced due to interplay between the interaction and commensuration. Whereas Ref.~\cite{BarnaliPRB} presents the expected features for weakly correlated Mott, but strongly correlated Mott exhibits stringent competition between Mott correlation and correlation due to disorder, which results to melting of Mott localization. 

In this paper,  we consider a few-boson ensemble in a finite 1D quasiperiodic lattice with various incommensurable setups. The motivation is to understand how the delocalization due to incommensurate fraction of particles in the periodic lattice competes to the localization introduced by disorder in the quasiperiodic lattice. The scenario is different whether the filling factor is less than or more than one. For filling factor less than one, the delocalization is justified due to poor population. For filling factor more than one, there is extra delocalized particles sitting on the commensurate background of localized particles. The challenging question is how the disorder introduced by the secondary incommensurate lattice will interplay with the existing delocalized fraction of particles in the primary lattice and determine the localization process.

We consider a typical setup of $S=7$ lattice sites of fixed lattice depth and different incommensurate setups are created by changing the number of bosons. However to unravel the key role played by the incommensurate filling, we also include the study for commensurate setup. We cover the different interaction regimes from the weakly correlated to the fermionization limit when bosons try to escape their spatial overlap. We go beyond the Bose-Hubbard model which assumes unperturbed Wannier orbitals and is restricted to the lowest band. However, one needs to include higher band effects to examine the fermionization and strong correlation effects. As the present work going to explain the underlying microscopic mechanism in the localization process and there is no new emergent phases due to incommensurate filling, we dedicatedly omit the phase diagram which is mostly understood for bulk system. This is also to mention that few-particle systems can be produced deterministically in the experiments and facilitates the investigation of fundamental building blocks of many-body systems from a bottom-up perspective~\cite{RevModPhys.80.885,RevModPhys.83.863}. Few atoms in the strongly interacting limit are highly correlated and can be handled numerically with high precision. The observed study can be taken as 'finite-size-precursor' of macroscopic phases in the thermodynamic limit. 

We investigate the ground state properties from a bottom-up {\it ab initio} perspective using multiconfigurational time dependent Hartree for bosons (MCTDHB), which is numerically exact and is an efficient technique for treating the interacting bosons specially in the strong interaction regime when the interatomic correlation plays a significant role~\cite{Sascha:2010,Meyer:2000,Fischer:2024}. We explore the ground states as a function of increasing interaction strength for different filling factor. We analyze the one- and two-body densities, fragmentation to understand the complex interplay between disorder and existing delocalization in the primary lattice.  

We consider the following set ups in the periodic lattice and further investigate their response in the presence of secondary lattice. a) commensurate filling, filling factor $\nu$= $\frac{N}{S}$ =1, $N=7$ bosons are in $S=7$ lattice sites. For weak interaction, we retrieve the usual localization process, although find robustness of strongly correlated Mott in the fermionization limit.  b) Incommensurate filling factor $\nu < 1$, $N=5$ bosons are in $S=7$ lattice sites. For weak interaction, although very strong disorder introduces central localization, some unseen features is observed in the strongly interacting limit. We observe that disorder drags the fragmented superfluid of the primary lattice to the fully fragmented Mott insulating phase.  c)  Incommensurate filling $\nu >1$ exhibits more intriguing competition between delocalization due to repulsion, on-site interaction and localization due to disorder. There is always $N$ ${\rm mod}$ $S$ particles at the background of Mott localization with unit filling. The extra particle/particles lie in the Mott background, in the fermionization limit they form one or more dimers. The introduced disorder displaces the dimers' position when the background Mott correlation does not change significantly.  

The paper is organized as follows. In Sec II, we present the theoretical framework behind the system we study. In Sec III,  we give a brief review of the method we employ to investigate the system numerically exactly. In this section we also define the observables used to understand different features.  In Sec. IV, we present results for commensurate filling. In Sec. V, we present results for incommensurate filling distributed over two subsections. We conclude our findings in Sec. VI.

\section{System and Protocol}
We study the ground state of $N$ interacting bosons of mass $m$ in a one-dimensional quasiperiodic lattice.
The system is described by the time-dependent many-body Schr\"odinger equation 
\begin{equation}
\hat{H} \Psi = i \hbar \frac{\partial \Psi}{\partial t}
\label{eq:TISE}
\end{equation}
The many-body Hamiltonian is written as
\begin{equation} 
\hat{H}(x_1,x_2, \dots x_N)= \sum_{i=1}^{N} \hat{h}(x_i) + \sum_{i<j=1}^{N}\hat{W}(x_i - x_j).
\label{eq:Hamiltonian}
\end{equation}
The one-body part of the Hamiltonian is $\hat{h}(x) = \hat{T}(x) + \hat{V}_{OL}(x)$, where $\hat{T}(x) = -\frac{\hbar^2}{2m} \frac{\partial^2}{\partial x^2}$ is the kinetic energy operator and $\hat{V}_{OL}(x) = V_p \sin^2(k_p x) + V_d \sin^2(k_d x)$ is an external potential characterized by two interfering standing waves that generate a quasiperiodic lattice.

The quasiperiodic potential consists of a primary lattice of depth $V_p$ and wave vector $k_p$ perturbed by a detuning lattice of depth $V_d$, wave vector $k_d$.
When the ratio of the two wave vectors $k_p/k_d$ is chosen to be incommensurate, the external potential generates a regular but non-repeating structure that implements correlated disorder on top of the periodic primary lattice.
The Hamiltonian in Eq.~\eqref{eq:Hamiltonian} can be written in dimensionless units obtained by dividing the dimensionful Hamiltonian by $\frac{\hbar^2}{m\bar{L}^2}$, with $\bar{L}$ an arbitrary length scale.
For the sake of our calculations, we will set $\bar{L}$ to be the period of the primary optical lattice.
Unless otherwise stated, we will keep the wave vectors fixed at $k_p=1.0$ and $k_d=1.19721$. For the entire calculation we fix $V_p=10 E_r$ and probe various values of the potential depths $V_d \in [0 E_r, 3 E_r]$, where $E_r = \frac{\hbar^2 k_p^2}{2 m}$ is the recoil energy of the primary lattice.
In the entire computation, we restrict the system geometry to accommodate $S=7$ sites in the primary optical lattice (defined as the spatial extent between two maxima in the sinusoidal function) adding hard-wall boundary conditions at each end. The typical case with seven wells, will well capture all essential features of competition between the attempted localization due to disorder and existing delocalization due to incommensuration. To probe the physics in commensurate filling we choose $N=7$ particles, whereas for incommensurate filling $<1$, $N=5$ and for $>1$, $N=8$ and $N=9$ are used.

The two-body interactions $\hat{W}(x_i - x_j)$ appearing in Eq.~\eqref{eq:Hamiltonian} are modeled as a contact interaction $\hat{W}(x_i - x_j)=\lambda \sum_{i<j}\delta(x_i-x_j)$. $\lambda$ controls the strength of interaction and it is positive (repulsive interaction) in our entire calculation. To probe the pathway from superfluid to fragmented SF to Mott and fully fragmented Mott, we probe the interaction strength in the range $\lambda \in [ 0 E_r, 10 E_r]$

We will solve the Hamiltonian above numerically exactly to extract the ground state properties of the system and map out relevant observables such as, real-space density, and correlations to reveal the fate of the many-body state subject to quasiperiodicity.

\section{Method} 
To study the ground-state properties of interacting bosons with correlated quasiperiodic disorder we employ the MultiConfigurational Time-Dependent Hartree method for indistinguishable particles (MCTDH)~\cite{Streltsov:2006, Streltsov:2007, Alon:2007, Alon:2008,Lode:2016,Fasshauer:2016} as implemented 
in MCTDH-X package~\cite{Lin:2020,MCTDHX}. 

In the MCTDH method, the many-body Schrödinger equation is represented by  the many-body wave function as an adaptive, time-dependent superposition of permanents, which are built from \( M \) single-particle wave functions known as orbitals. 
In this approach both the coefficients and the basis functions are optimized over time which enables the calculation of ground-state properties through imaginary relaxation and the simulation of full-time dynamics using real-time propagation.
In this work, we focus exclusively on imaginary time propagation to relax the system to its ground state.
Further details on this method are contained in Appendix A.

Utilizing several orbitals in the many-body ansatz allows us to capture the beyond mean-field scenario when the many-body state is fragmented- when several orbitals have significant occupation. The convergence of the method is guaranteed in the limit $M \to \infty$. 
However in practice, a much smaller set of orbitals is needed to obtain \emph{numerically exact} results. For strongly interacting systems, a larger number of orbitals might be required to capture many-body correlation effects. Computation is repeated by increasing the number of orbitals and optimized when the results do not change by including more orbitals in the MCTDH expansion. In our entire calculation we use $M=12$ orbitals. 

To study the ground state of the system, we compute various observables from the many-body state $\left| \Psi \right>$.
  To get the information about the spatial distribution of the bosons we calculate the one-body density as 
\begin{equation}
\rho(x)=\langle\psi(x)|\hat{\psi}^\dag(x)\hat{\psi}(x)|\psi(x)\rangle.
\end{equation}
To measure the degree of coherence in the many-body correlation we calculate the reduced one-body and two-body density matrices defined as 
\begin{align}
\rho^{(1)}(x,x') &= \left< \Psi \right| \hat{\Psi}^{\dagger}(x) \hat{\Psi}(x') \left| \Psi \right> \\
\rho^{(2)}(x,x') &= \left< \Psi \right| \hat{\Psi}^{\dagger}(x) \hat{\Psi}^{\dagger}(x') \hat{\Psi}(x') \hat{\Psi}(x) \left| \Psi \right>.
\end{align}

Information about the orbital occupations in the ground state can be extracted from reduced one-body density matrix.
This is described in terms of the natural orbitals $\phi_i^{(\mathrm{NO})}$ and their occupations $\rho_i$, which correspond to the eigenfunctions and eigenvalues of $\rho^{(1)}(x,x')$, respectively, as given by
\begin{equation}
\rho^{(1)}(\mathbf{x},\mathbf{x}') = \sum_i \rho_i \phi^{(\mathrm{NO}),*}_i(\mathbf{x}')\phi^{(\mathrm{NO})}_i(\mathbf{x}).
\label{eq:RDM1}
\end{equation}
We further utilize the eigenvalues of the reduced one-body density matrix to construct an order parameter~\cite{budha1}
\begin{equation}
\Delta = \sum_i \left( \frac{\rho_i}{N} \right)^2
\label{eq:order-par}
\end{equation}
$\Delta$ quantifies the loss of coherence and thus maps out the transition from superfluid ($\Delta \to 1$) to fragmented Mott state ($\Delta \to \frac{1}{S}$).

The one-body momentum distribution is constructed from 
\begin{equation}
    n(k) = \int dx \int dx^{'} e^{-ik(x-x^{'})} \rho^{(1)}(x,x')
\end{equation}

\section{Localization and correlation in disordered lattice for commensurate filling}
\begin{figure}[t]
    \centering
         \includegraphics[width=0.7\textwidth, angle=270]{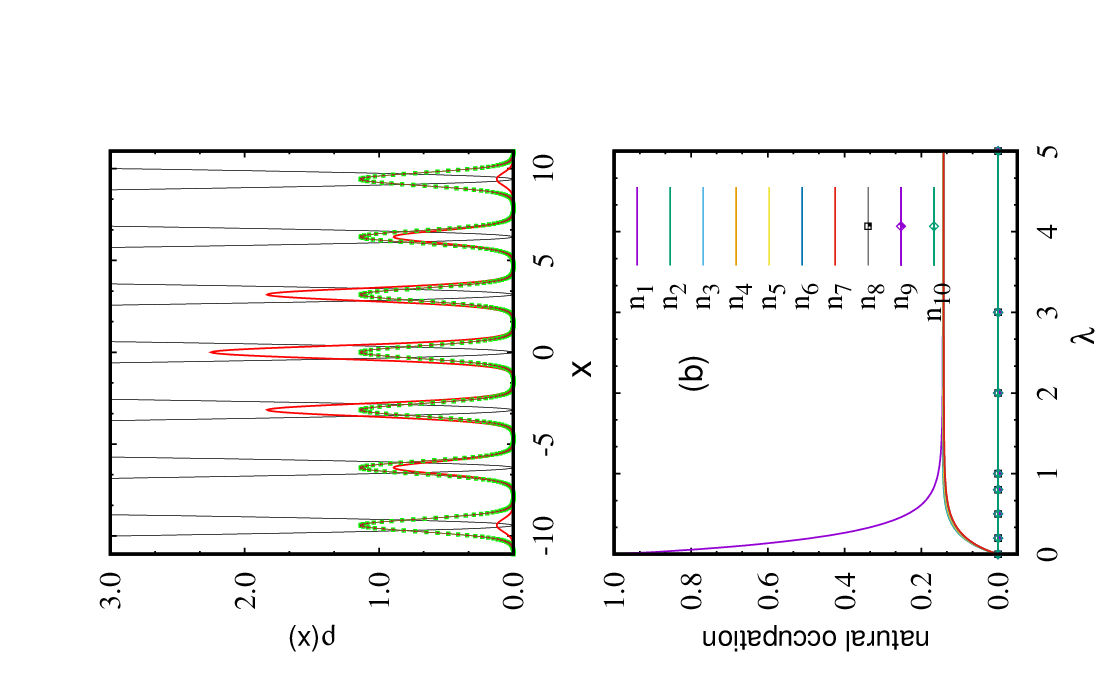}
    \caption{(a)The one-body density $\rho(x)$ in the primary lattice with seven wells and seven particles for different values of interaction strength. Solid red line, lines with green solid square and brown down triangle correspond to $\lambda=0 E_r, 0.5 E_r, 5 E_r$ respectively. The black line represents the primary lattice. For $\lambda=0 E_r$, the density is maximal at the central well- superfluid phase. For $\lambda=0.5 E_r$, the bosons are equally distributed in seven wells- Mott-insulator phase.  For higher interaction strength, $\lambda=5 E_r$, one-body density overlaps with the density for $\lambda=0.5 E_r$- Mott localization is maintained. (b) Populations of the first ten natural orbitals as a function of interaction strength. As $\lambda$ increases, the occupation of the first orbital gradually decreases while other orbitals begin to contribute. Eventually the many-body state becomes seven-fold fragmented ($n_1 \simeq n_2 \simeq n_3 \simeq n_4 \simeq n_5 \simeq n_6 \simeq n_7 \simeq 1/7$ at $\lambda \simeq 3$. The remaining three orbitals have insignificant contributions in the entire range of $\lambda$.}
    \label{fig1}
\end{figure}

In this section, we present ground state properties in the periodic and quasiperiodic lattice for the commensurate filling of $N=7$ bosons in $S=7$ lattice sites for various interaction strength, ranging from delocalized superfluid in the noninteracting limit to fragmented Mott insulating phase in the fermionization limit. The one-body is presented in Fig.~\ref{fig1}(a). 
In an optical lattice, the competition between the kinetic, trap and interaction energy determines the ground state configuration. For the noninteracting case, the kinetic and trap energy dominates, the density is maximal in the central well representing a superfluid phase. However as the interaction strength increases, there will be redistribution of the density in different sites. 
For $\lambda=0.5 E_r$, bosons localize equally in all sites. This Mott-insulator phase is characterized by the localization of atoms in the lattice with a vanishing overlap of the densities in distinct wells. Once localization occurs, the increase in the interaction ($\lambda=5 E_r)$ will no longer affect the density distribution. 

The superfluid-to-Mott-insulator transition is further characterized by fragmentation. In Fig.~\ref{fig1}(b), we plot the corresponding occupation of the first $M=10$ orbitals as a function of the interaction strength. We use the criterion of Penrose and Onsager~\cite{Onsager} to identify whether the many-body state is condensed or fragmented. The system is a condensate if the reduced one-body density matrix has a single macroscopic eigenvalue. For noninteracting case or very very weak interaction, when the first natural orbital is macroscopically occupied and the population of the remaining orbitals is insignificant, the system can be well approximated by the many-body wave function with a single-orbital mean-field state $\left |N, 0.....\right > $. With increase in $\lambda$, the occupation of the first natural orbital gradually decreases while the occupations of other orbital gradually increase. For $\lambda=0.5 E_r$, when the one-body density exhibits equal population in all sites, the many-body state is a fragmented Mott state characterized by significant occupations in several orbitals. Although completely fragmented  Mott is not achieved. Fully fragmented Mott is achieved when the number of contributing orbitals becomes equal to the number of lattice sites and each orbital contributes equally, $\frac{1}{S}$ of total population. With further increase in $\lambda$, the Mott insulating phase gradually drives to fully fragmented Mott phase- the many-body wave function is a seven-fold fragmented state $ \left | 1,1,1,1,1,1,1 \right>$, each orbital contributes $\simeq 14.28 \%$. Fig.~\ref{fig1} demonstrates the transition from SF to fragmented MI phase. 

\begin{figure}
    \centering
         \includegraphics[width=0.4\textwidth, angle=270]{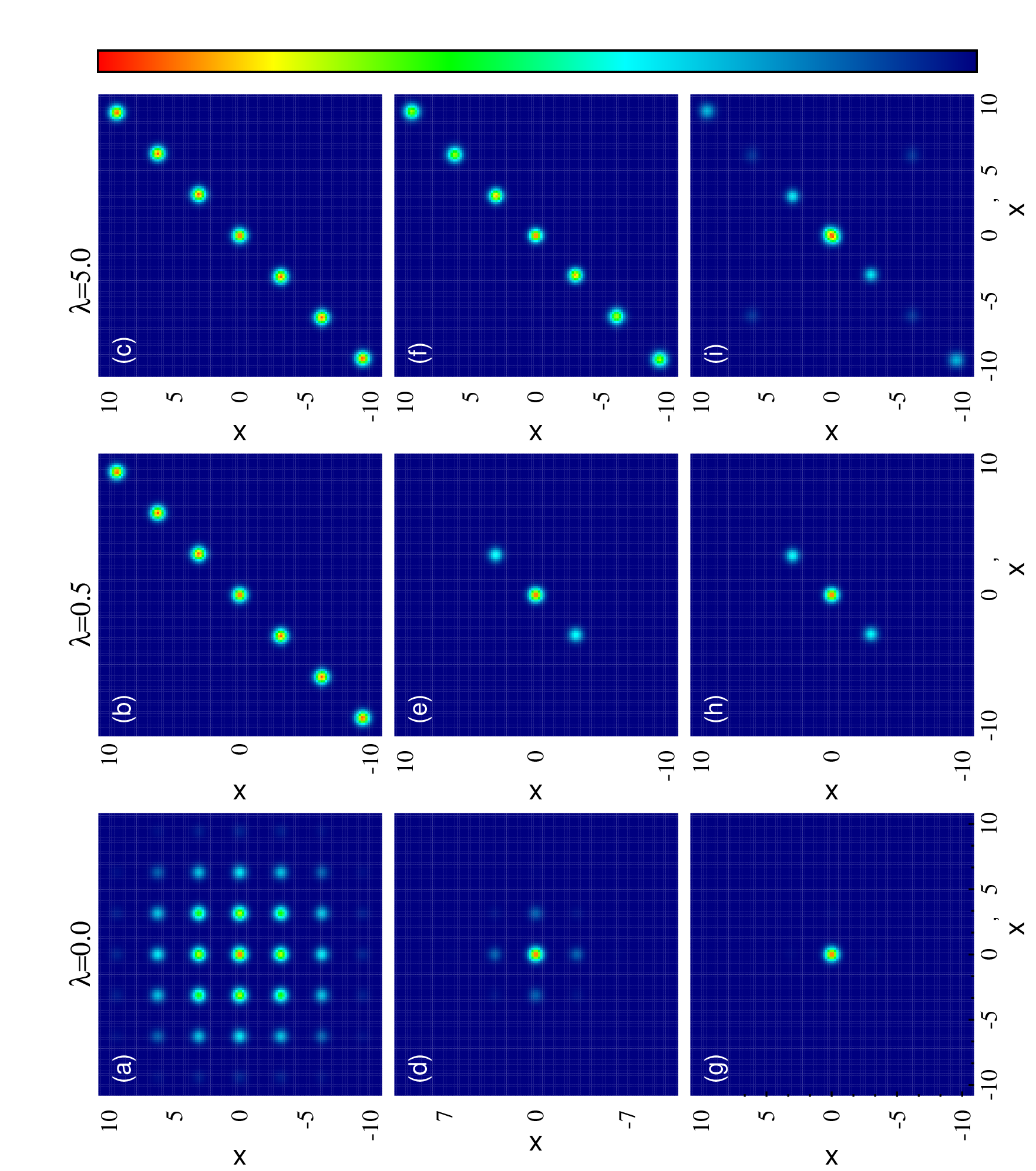}
    \caption{The reduced one-body density $\rho^{(1)}(x,x^{\prime})$ for commensurate filling. The panels show a comparison between periodic [$V_d=0 E_r$, panels (a)-(c)] and various choices of quasiperiodicity [panels (d)-(i)]. For $\lambda=0 E_r$, (d): $V_d=0.01 E_r$ and (g): $V_d=0.05 E_r$. For $\lambda=0.5 E_r$, (e): $V_d=1 E_r$ and (h): $V_d=3 E_r$. For $\lambda=5 E_r$, (f): $V_d=2 E_r$ and (i): $V_d=3 E_r$. }
    \label{fig2}
\end{figure}

To illustrate how the correlated disorder modifies the density signature in the primary lattice, we further study the one-body reduced density matrix $\rho^{(1)}(x,x')$ for several values of $V_d$. We choose three specific cases of interaction strength.; $\lambda=0.0$ (superfluid), $\lambda=0.5$ (Mott-insulating), $\lambda=5.0$ (fragmented Mott) [Figs.~\ref{fig2}(a)-(i)]. In the quasiperiodic lattice, the response of Mott insulating phase and fully fragmented Mott phase will be different, the later remains robust even for strong disorder. 

For noninteracting case ($\lambda=0 E_r$) and in the periodic lattice ($V_d=0 E_r$), the system is a delocalized superfluid. The short and long-range correlation are manifested in the non-vanishing off-diagonal behavior of $\rho^{(1)}(x, x')$ ((Fig.~\ref{fig2}(a)). Although for such a finite ensemble true off-diagonal long-range order is absent, however off-diagonal of $\rho^{(1)}(x, x')$ measures correlation across the lattice. We also observe the off-diagonal correlation is gradually depleted with increasing distance from the central lattice. On switching on small disorder ($V_d=0.01 E_r$) immediately destroys the off-diagonal correlation (Fig.~\ref{fig2}(d)) and $V_d=0.05 E_r$ leads to localization in the central lattice (Fig.~\ref{fig2}(g)) as expected.

When $\lambda$ increases, the off-diagonal correlation is already depleted in the periodic lattice (Fig.~\ref{fig2}(b)). Weak repulsion ($\lambda=0.5 E_r$) pulls the  particles away from the middle wells, Mott localization happens. Seven particles are uniformly distributed in seven lattice sites and only the intra-well correlation is maintained whereas inter-well correlation is completely lost. The diagonal of $\rho^{(1)}(x, x')$ shows seven completely separated coherent regions where $\rho^{(1)} \simeq 1$ and at the off-diagonal $\rho^{(1)} \simeq 0$ (Fig.~\ref{fig2}(b)). The Mott localization is maintained even for weak disorder. For sufficiently strong disorder ($V_d=1 E_r$) the Mott localization is gradually destroyed which further leads to localization into middle and two side wells (Fig.~\ref{fig2}(e)). However, even very strong disorder ($V_d=3 E_r$), fails to bring central localization (Fig.~\ref{fig2}(h)).  

The situation is more interesting in the strongly interacting limit $\lambda=5 E_r$, when the fully fragmented Mott insulating phase in the primary lattice is strongly correlated. Although it is impossible to distinguish the Mott correlation for weak interaction (Fig.~\ref{fig2}(b)) and strong interaction (Fig.~\ref{fig2}(c)), as in both cases the Mott localization is only manifested by seven bright lobes along the diagonal. However their response in quasiperiodic lattice is different. The less correlated Mott is dragged to localization in the quasiperiodic lattice, whereas strongly correlated Mott remains robust even when the disorder is quite high $V_d=2 E_r$ (Fig.~\ref{fig2}(f)). For much stronger disorder, $V_d=3 E_r$ the Mott correlation in the outer lattice sites is slightly depleted but localization in the middle wells does not happen (Fig.~\ref{fig2}(i)).

%

\section{Localization and correlation in disordered lattice for incommensurate filling}

In this section we like to investigate the consequences of disorder for different incommensurate filling factors both smaller and greater than one. The competition between localization due to stronger interaction and localization due to disorder becomes more fascinating for incommensurate filling due to additional existing delocalization from superfluid fraction. For filling factor less than one, due to less population of particles, on-site interaction does not dominate. In the noninteracting case, disorder introduces the central localization as expected. However in the limit of very strong interaction, particles are already driven away from the central lattice, the introduced disorder by the secondary lattice tries to localize them in the middle lattice sites. We observe that the localized many-body state exhibits the signature of Mott localization.  Filling factor more than one, combines localization, delocalization as well as on-site interaction. Due to extra population of particles, on-site effect is manifested leading to beyond Bose-Hubbard physics. In the fermionization limit, the lattice site with double occupation forms dimer and the pair fragments into two orbitals in the same well. The additional disorder now competes the situation in a very complex way. In both cases of incommensurate filling we present the key measures - one and two-body correlation functions.

\subsection{Filling factor $\nu < 1$}

\begin{figure}[t]
    \centering
         \includegraphics[width=0.5\textwidth, angle=270]{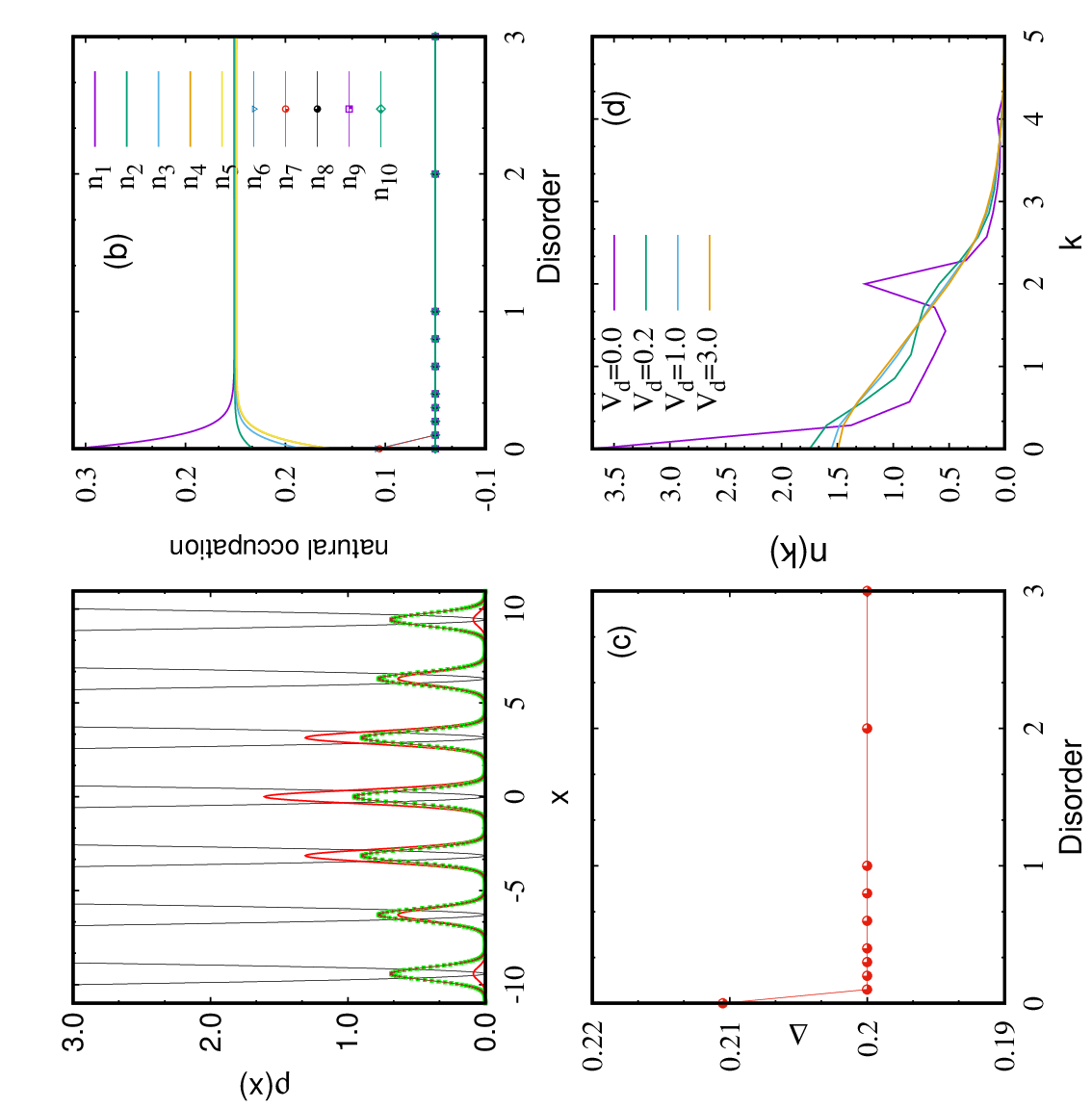}
    \caption{(a) One-body density $\rho(x)$ in the primary lattice with five bosons in seven lattice sites and for different interaction strength. Solid red line, lines with green solid square and brown down triangle correspond to $\lambda=0 E_r, 0.5 E_r, 5 E_r$ respectively. The black line represents the primary lattice. (b) Natural occupations in the lowest $M=10$ orbitals for $\lambda=5 E_r$ as a function of disorder. (c) Order parameter for $\lambda= 5 E_r$ as a function of disorder. (d) Momentum distribution for $\lambda=5 E_r$ and for various disorder parameter $V_d$.}
    \label{fig3}
\end{figure}

In this section, we consider the case of incommensurate filling with $\nu <1$, five bosons in seven wells. Fig.~\ref{fig3} summarizes our main observation, how the particles localize in the lattice in the presence of disorder. Fig.~\ref{fig1}(a) represents the ground-state density in the primary lattice for various interaction strength $\lambda=0 E_r, 0.5E_r, 5 E_r$. For the noninteracting ground state, the density is larger in the middle sites and decreases as we go to the outer site. Now the main concern how the particles will distribute themselves in different sites with increase in interaction strength. In the weakly interacting limit, $\lambda=0.5 E_r$, there is a transfer of particles from the middle wells to the outer wells. As $N < S$, one-particle density does not tend to an equal site occupation even in the strongly interacting limit $\lambda=5 E_r$. 

To understand the response of the system in the presence of secondary lattice we utilize several key measures as a function of disorder strength : fragmentation, order parameter, one-body momentum distribution, one-body and two-body correlation function. For the noninteracting case, disorder simply introduces localization in the central lattice as expected and not shown here. The rest of the analysis in this section is focused for strongly interacting limit, $\lambda=5 E_r$. In Fig.~\ref{fig3}(b), we present the fragmentation analysis of natural orbitals. The relative population $n_i$ in different natural orbitals defines the degree of fragmentation. For a nonfragmented condensate the highest occupation in the lowest natural orbital $\simeq 1$. Fragmentation happens when more than one natural orbital exhibit significant population. The computation is done with $M=12$ orbitals and in Fig.~\ref{fig3}(b), we present normalized population in lowest $M=10$ natural orbitals (as the population in the other two orbitals is $<$ $10^{-6}$) as a function of the disorder strength $V_d$. When a single eigenvalue is macroscopic, the state is a superfluid. However, for $V_d=0 E_r$, the ground state is already fragmented, as several orbitals contribute, it is a fragmented superfluid. With increase in $V_d$, the occupation of the first natural orbital gradually decreases while the occupations of the second, third, fourth and fifth orbitals gradually increase. For $V_d=0.5 E_r$, the first five natural occupations equally saturate at $20 \%$, all other remaining orbitals had occupations smaller than $10^{-6}$, thus leading to five-fold fragmentation. This five-fold fragmentation indicates the transition to the five-fold fragmented insulating phase which persists even for very strong disorder. Thus introduced disorder drags the fragmented superfluid to a insulating phase localized in five middle sites. Generally, a fully fragmented Mott-insulating phase is characterized when there are as many significant eigenvalues of the reduced density matrix as there are lattice sites. Thus strongly interacting fragmented superfluid of $N=5$ bosons in the primary lattice settles to fully-fragmented localized insulating phase in the $S=5$ middle sites in the presence of secondary lattice.  The localized phase can be called as five-fold fragmented Mott phase. The transition can be more clearly inferred from the one- and two-body correlation functions as discussed later. 

In Fig.~\ref{fig3}(c), we show the order parameter $\bigtriangleup$, as a function of detuning lattice depth $V_d$. When superfluid exhibits a single orbital occupation, typically $\bigtriangleup \simeq 1$. As Mott insulating phase exhibits equally distributed density over multiple orbitals and sites $\bigtriangleup \simeq \frac{1}{S}$. Fig.~\ref{fig3}(c) exhibits that in the primary lattice, the initial state is already fragmented due to strong interaction. With increase in $V_d$, the insulating phase emerges when $\bigtriangleup$ reaches to $\frac{1}{S} =0.2$, i.e., the five-fold fragmentation happens. For weaker interaction $\lambda=0.5$ (not shown here), the initial state is less fragmented and gradually reaches to fully fragmented insulating phase with increase in detuning lattice depth. 

In Fig.~\ref{fig3}(d), we observe that the momentum distribution exposes a rich pattern with Bragg peaks for $V_d=0 E_r$. The central peak is also high signifying off-diagonal long-range order. With increase in $V_d$, the Bragg peak is gradually smeared, the central peak is substantially lowered which is a typical signature of loss of correlation due to localization. 

\begin{figure}[t]
    \centering
         \includegraphics[width=0.6\textwidth, angle=270]{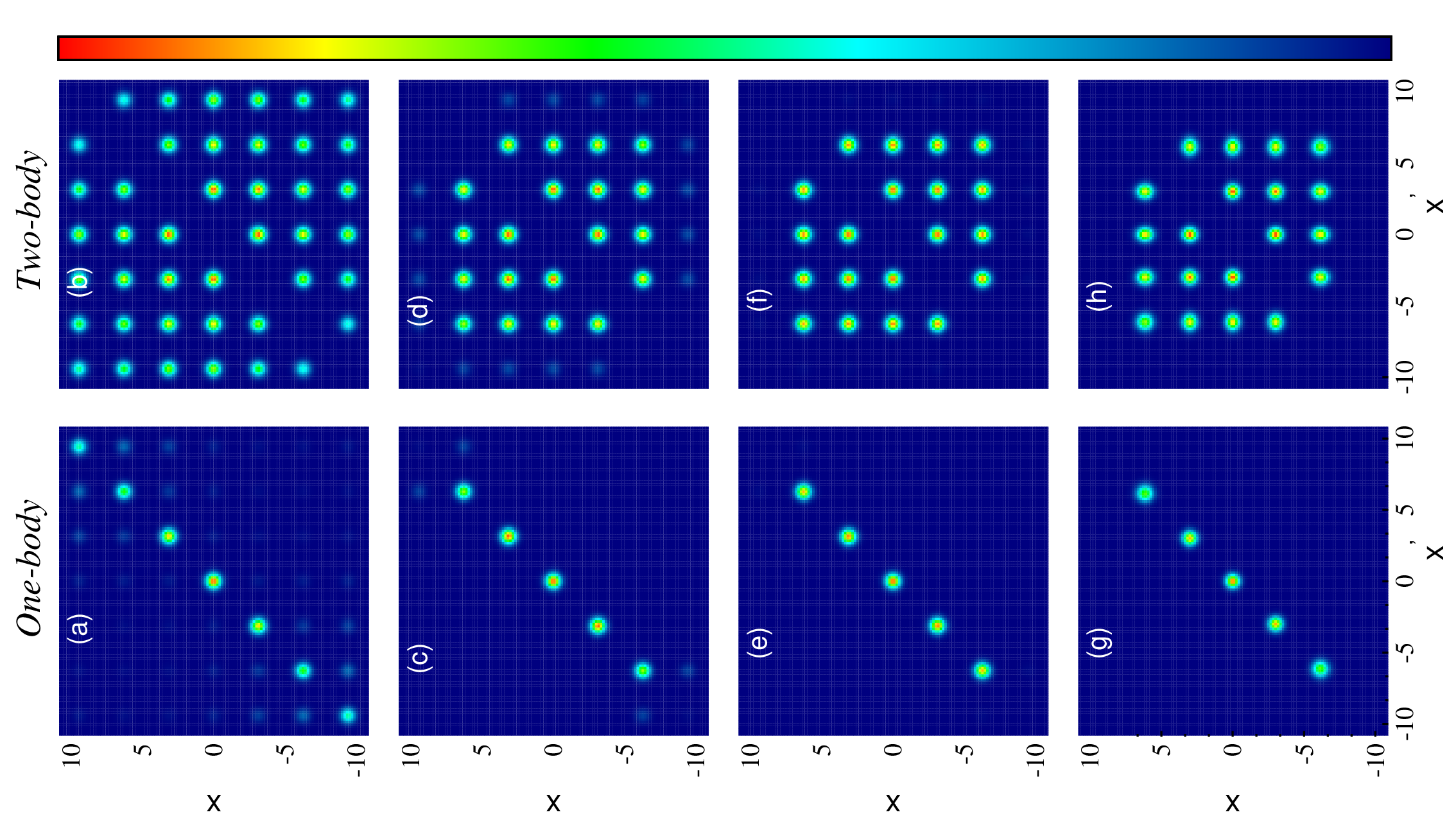}
    \caption{The reduced one-body density $\rho^{(1)}(x,x^{\prime})$ (left column) and two-body density $\rho^{(2)}(x,x^{\prime})$ (right column) for five bosons in seven wells with interaction strength $\lambda=5 E_r$. The panels show comparison between periodic case [$V_d=0 E_r, (a)-(b)$] and the quasiperiodic cases [panels (c)-(h)]. $V_d=0.1 E_r$ [(c)-(d)], $V_d=0.5 E_r$ [(e)-(f)], $V_d=3 E_r$ [(g)-(h)]}
    \label{fig4}
\end{figure}

In Fig.~\ref{fig4}, we present both the one- and two body correlations for the incommensurate setup with five bosons in seven wells in the strongly interacting limit with $\lambda=5E_r$. The left column corresponds to one-body and the right column corresponds to two-body correlation for various de-tuning lattice depth $V_d$. The main difference between the commensurate and incommensurate filling, is the off diagonal part of reduced one-body density matrix can not vanish completely due to the presence of delocalized superfluid. In the clean lattice, the one-body correlation is mostly concentrated along the diagonal, however not equally distributed in all sites. There is a quite well localization around the middle lattice sites, with vanishing off-diagonal correlation. Whereas in the distant lattice sites, localization is poor which is also confirmed by the presence of off-diagonal correlation (Fig.~\ref{fig4}(a)). The two-body correlation in the clean lattice shows an almost depleted diagonal implying $\rho^{(2)}(x,x) \rightarrow 0$, probability of finding two bosons in the same lattice is zero. The square-lattice like pattern with  a missing diagonal is the hallmark of insulating phase. However in Fig.~\ref{fig4}(b), for $V_d=0 E_r$, the two-body off-diagonal coherence is not uniform signifying the nonuniform localization of the particles and presence of delocalized superfluid fraction. On switching on the disorder potential, $V_d=0.1E_r$, it immediately dominates and eliminates the existing delocalization of the primary lattice (Fig.~\ref{fig4}(c)-(d)). For $V_d=0.5 E_r$, five bosons localize in the five central lattice sites. Five fully coherent bright spots along the diagonal signifies Mott localization happens in the five middle sites while two outer lattice sites remain empty (Fig.~\ref{fig4}(e)). The two-body correlation function with completely extinguished diagonal and uniform off-diagonal correlation also settles around the five middle lattice sites (Fig.~\ref{fig4}(f)), thus confirming Mott localization. The feature remains same even for very strong disorder $V_d=3 E_r$ (Fig.~\ref{fig4}(g)-(h)).    

\subsection{Filling factor $\nu >1$}

For a filling factor more than one, when $N$ mod $S$ number of particles always reside at the background of commensurate filling, the on-site interaction interfere the localization process in the presence of secondary lattice. We observe strong on-site repulsion forms dimer structure in the densities. We consider two specific cases: eight particles in seven lattice i.e, one extra particle added to the unit filling case and nine particles in seven lattice i.e., a pair of repulsive particles sits on a unit filling background. In the presence of disordered lattice; localization, delocalization and on-site interaction interplay in a complex way. 

\begin{figure}[t]
    \centering
         \includegraphics[width=0.7\textwidth, angle=270]{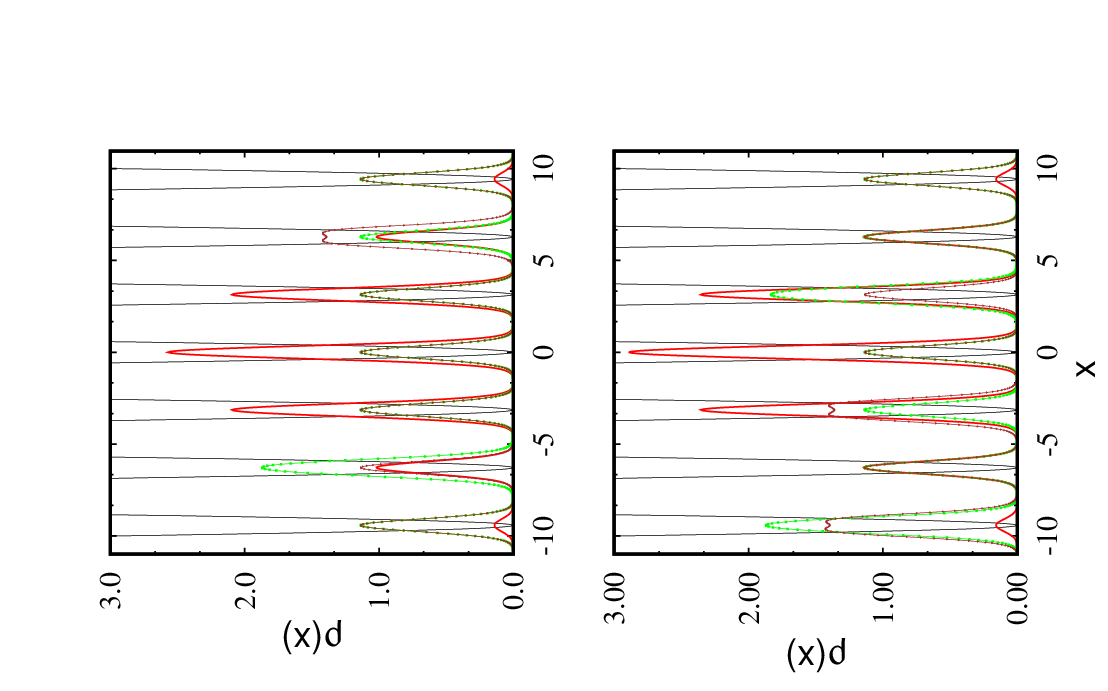}
    \caption{(Top panel) One-body density $\rho(x)$ in the primary lattice with eight bosons in seven lattice sites and for different interaction strength. Solid red line, lines with green solid square and brown down triangle correspond to $\lambda=0 E_r, 3 E_r, 10 E_r$ respectively. (Bottom panel) One-body density $\rho(x)$ in the primary lattice with nine bosons in seven lattice sites and for different interaction strength. Solid red line, lines with green solid square and brown down triangle correspond to $\lambda=0 E_r, 3 E_r, 10 E_r$ respectively. The black line represents the primary lattice. }
    \label{fig5}
\end{figure}

\subsubsection{One extra particle on localized background}
The top panel of Fig.~\ref{fig5} represents the ground state density in the clean lattice for eight particles distributed over seven lattice sites. For noninteracting case, the density is non uniform and centrally localized. For weak interaction, $\lambda=0.5 E_r$, the nonuniform occupation tends to become uniform (not shown here). For higher interaction, $\lambda=3 E_r$ (green curve), there is a predominant occupation in a side well, leading to nonuniform occupation. The high peak in the corner lattice signifies localization into pairs, the uniform Mott background exists. For much higher interaction, $\lambda=10 E_r$ (brown curve), the on-site repulsion is dominating. The onset of fermionization is mimicked by the formation of characteristic dip in the density distribution. In terms of fragmentation, the pair of interacting particles occupy two orbitals in the same well.  To understand the scenario, we need to go beyond Bose-Hubbard analysis when higher band contributions need to be taken into account. According to this, when seven particles occupy the lowest band forming an MI background of one particle localized in one well, the extra particle lies  on the energetically lowest one-particle level of excited band~\cite{Sascha:2010}. Thus, the extra one delocalized particle does not allow a perfect insulator phase. 

\begin{figure}[t]
    \centering
         \includegraphics[width=0.4\textwidth, angle=270]{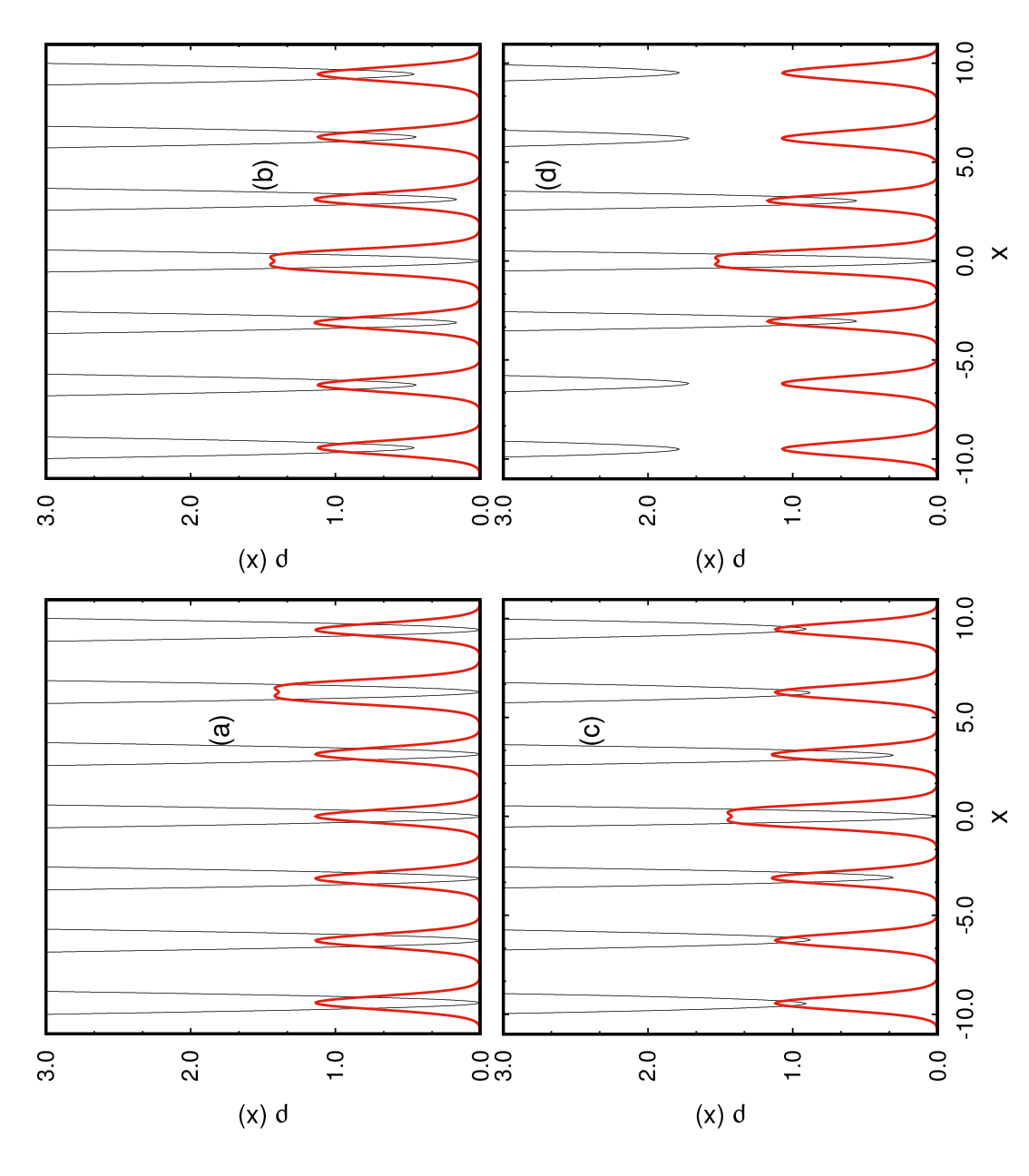}
    \caption{The one-body density $\rho(x)$ for eight bosons in seven sites for interaction strength $\lambda=10 E_r$ and for various quasiperiodic parameter: $V_d=0 E_r$ [(a)], $V_d=0.5 E_r$ [(b)], $V_d=1 E_r$ [(c)], $V_d=2 E_r$ [(d)]. For each case, the black dashed curve presents the corresponding disordered lattice potential.}
    \label{fig6}
\end{figure}

In Figs.~\ref{fig6}(a)-(d), we present the one-body density in the fermionization limit $\lambda=10 E_r$ in presence of disorder, $V_d=0 E_r,0.5 E_r,1 E_r,2 E_r$ respectively. The pair of interacting bosons relocates the lattice site in the disordered lattice, however, the commensurate background of the insulating phase with seven particles localized in the seven lattice sites remains unchanged. It is also to mention that for weaker interaction, when on-site effect is not predominating, the introduced disorder easily competes with the weakly correlated Mott background surrounded by superfluid layer leading to localization in the central wells (not shown here).

\begin{figure}[t]
    \centering
         \includegraphics[width=0.4\textwidth, angle=270]{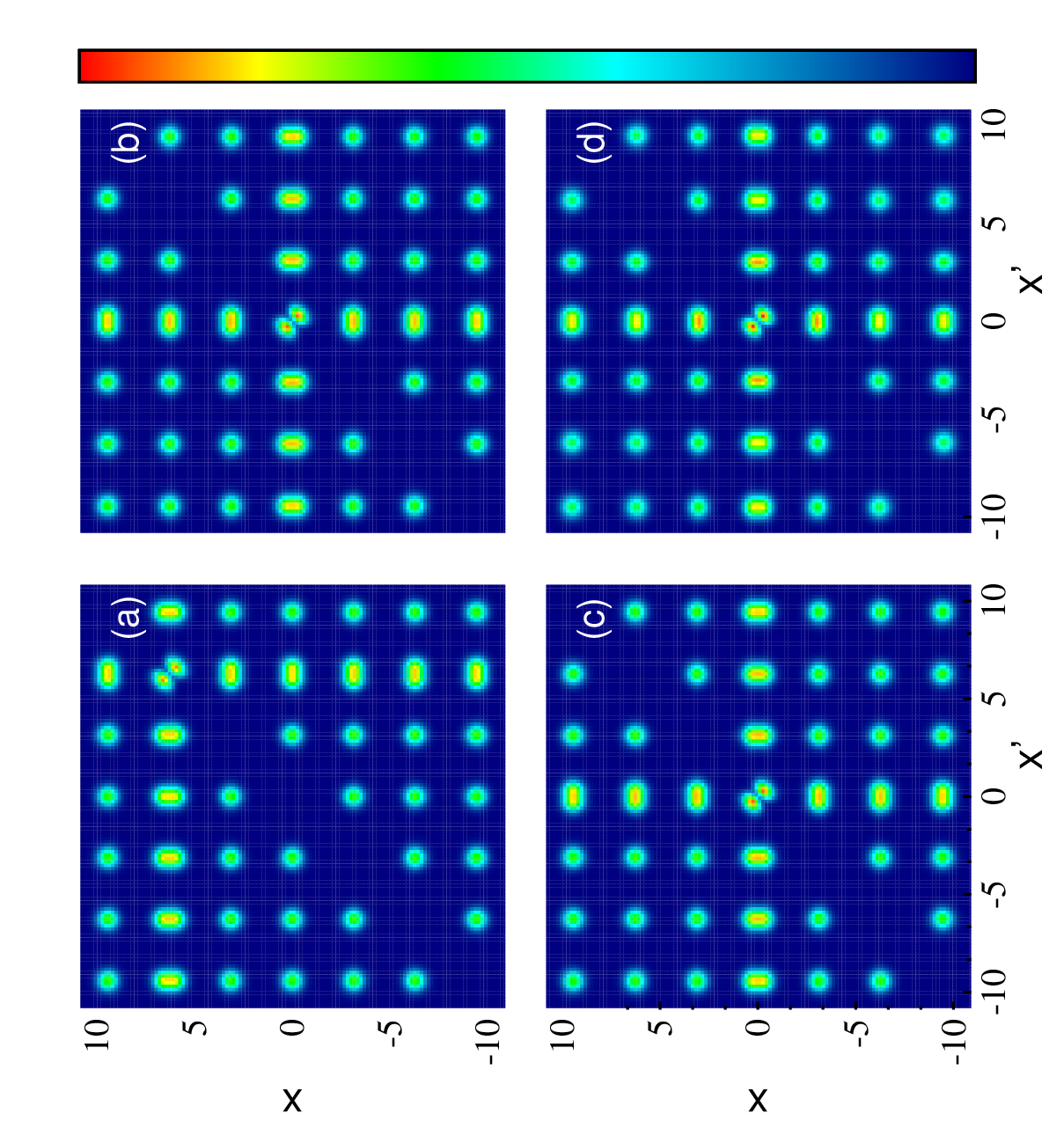}
    \caption{Behavior of the reduced two-body density $\rho^{(2)}(x,x^{\prime})$ for eight bosons in seven wells with interaction strength $\lambda= 10 E_r$. The panels show the comparison between the periodic [(a) $V_d=0 E_r$] and quasiperiodic cases
    [(b) $V_d=0.5 E_r$, (c) $V_d=1 E_r$, (d) $V_d=2 E_r$].
}
    \label{fig7}
\end{figure}

In Fig.~\ref{fig7}(a)-(d), we present the two-body density for periodic as well as for different disorder potentials in the fermionization limit. As stated before that completely depleted diagonal and the off-diagonal correlation is the hallmark of MI phase. In Fig.~\ref{fig7}(a), in the periodic lattice [$V_d=0 E_r$], Mott dimerization happens in a corner lattice. The depleted diagonal with a dimer formation exhibits that while seven particles completely occupy the lowest band states forming a Mott-insulator background, the extra particle lies in the energetically lowest level of the excited band. In quasiperiodic lattice $V_d=0.5 E_r$ (Fig.~\ref{fig7}(b)) the dimer is settled in the central lattice when the depleted diagonal remains unchanged signifying the Mott background does not loose any correlation. For higher disorder strength $V_d=1 E_r$ (Fig.~\ref{fig7}(c)) and  $V_d=2 E_r$ (Fig.~\ref{fig7}(d))the
position of the dimer does not change anymore. We also do not find any loss of Mott correlation in the range of correlated disorder we probe.

\subsubsection{Two extra particles on localized background}
Next we consider the case when a repulsive pair of particles resides on a localized background, nine particles settle in seven lattice sites. The bottom panel of Fig.~\ref{fig5} presents the one-body density $\rho(x)$ in the primary lattice for various interaction strength. The solid red line represent the noninteracting case where nonuniform distribution is displayed. In the stronger interaction, $\lambda= 3 E_r$ (green curve) represents two high peaks in two side lattices signifying localization of two pairs of bosons in each lattice site. Thus the many-body state is again a Mott dimerized state, but due to two extra particles, two dimers are formed in two sites. In the fermionization limit, $\lambda=10 E_r$, the characteristic dips appear, signifying that localized pairs are now fragmented. 

\begin{figure}[t]
    \centering
         \includegraphics[width=0.6\textwidth, angle=270]{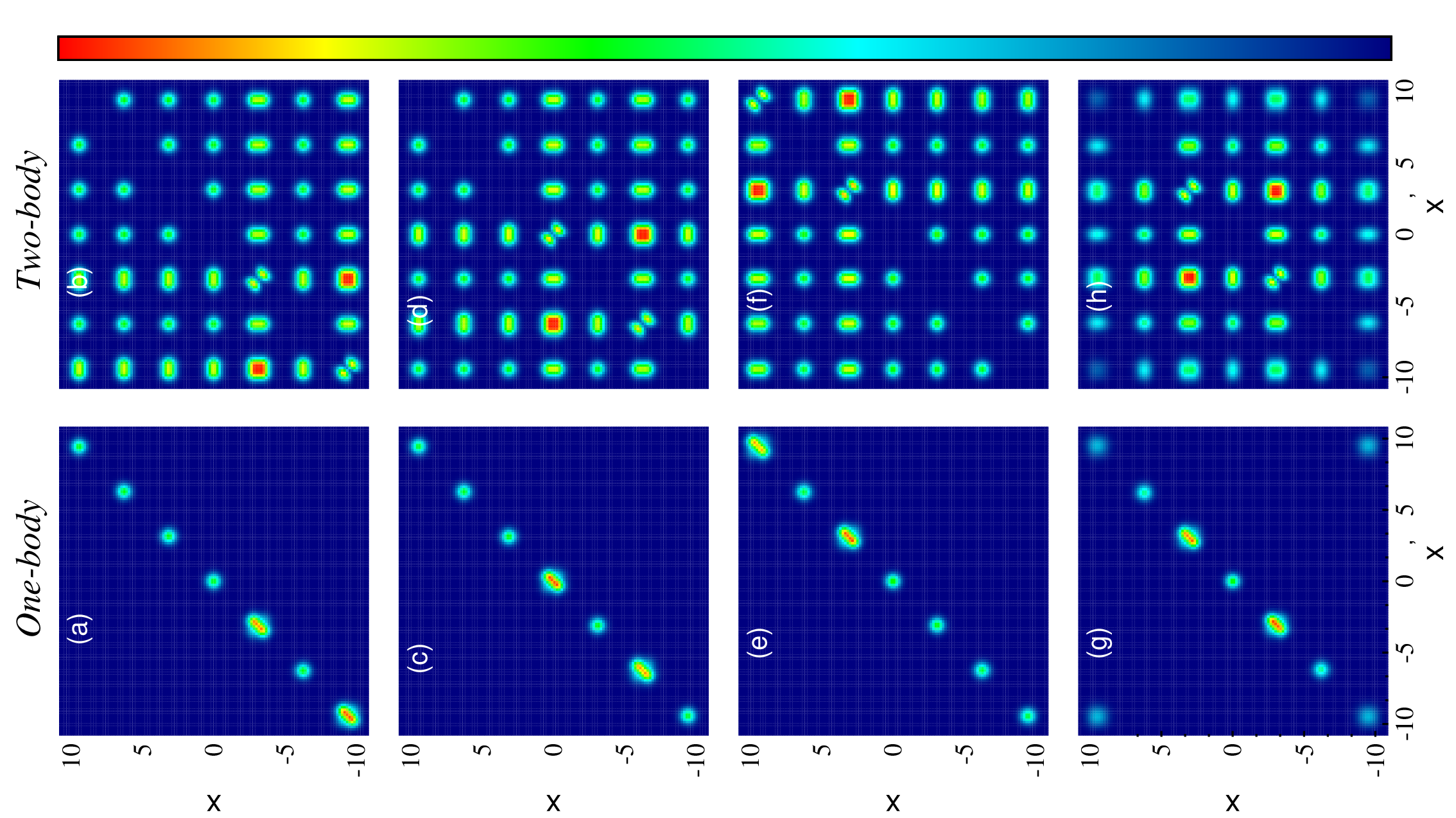}
    \caption{The reduced one-body density $\rho^{(1)}(x,x^{\prime})$ and two-body density $\rho^{(2)}(x,x^{\prime})$ for nine bosons in seven wells with interaction strength $\lambda=10 E_r$. The panels show comparison between periodic case [$V_d=0 E_r$, panels (a)-(b)] and the quasiperiodic cases [panels (c)-(h)]; $V_d=0.5 E_r$ [(c)-(d)], $V_d=1 E_r$ [(e)-(f)], $V_d=2 E_r$ [(g)-(h)]}
    \label{fig8}
\end{figure}

The one- and two-body correlation in the fermionization limit is presented in Fig.~\ref{fig8}. Some distinct features are expected as the two extra particles will now experience intra-pair repulsive force as well as repulsion due to the Mott background. For weaker interaction (not shown here) the off diagonal two-body correlation exists across the periodic lattice. Additional disorder leads to destroy the long-range correlation gradually. In the fermionization limit, $\lambda= 10 E_r$, the one-body correlation displays seven bright lobes along the diagonal and complete extinction of off-diagonal correlation. However, the lobes are not of uniform coherence, they are brighter in the two corner lattice sites signifying the localization of two extra particles (Fig.~\ref{fig8}(a)). In the two-body correlation function, the off-diagonal correlation is maintained and dimer structure with clear extinction of diagonal confirms the fragmented dimerization (Fig.~\ref{fig8}(b)). For $\lambda= 3 E_r$ (not shown here), we have observed the dimers appear but without the correlation hole signifying that the pairing bosons are not fragmented. By "corelation hole" we mean $\rho^{(2)}(x,x') \rightarrow 0$, implying that the probability of finding two bosons in the same place is reduced. This resembles the behavior of hard-core bosons with infinitely strong contact interactions. In the fermionization limit the dimers are more correlated due to combined effect of strong intra-pair interaction as well as strong interaction with the Mott background. Now weak disorder ($V_d=0.5E_r$) does not affect the background Mott correlation, only the dimer changes their location (Fig.~\ref{fig8}(c)-(d)). However for stronger disorders ($V_d=1 E_r$ and $2 E_r$), 
we observe reduction in Mott correlation as well as the dimers shift the lattice sites (Fig.~\ref{fig8}(e)-(h)). 

\section{Conclusion}

We have demonstrated the effect of disorder in a few-boson systems in finite 1D lattice for different filling factors and varying strength of the repulsive interactions. Our study reveals a rich interplay between quasiperiodic disorder and filling factor.  We have performed a numerically exact investigation  and the interplay between interaction, disorder and incommensuration is presented by the measures of fragmentation, one- and two-body correlation. For the commensurate filling, we observe the expected physics: disorder introduced localization happens in noninteracting and very weakly interacting superfluid. Whereas, for strong repulsive interaction, the strongly correlated fragmented Mott localization remains robust, maintaining its structural integrity even under strong disorder strength. 

However for incommensurate filling our study uncovers more intriguing interplay between disorder and interaction strength. It unravels some features not observed for commensurate filling. For the incommensurate filling in the primary lattice, the initial state is partially delocalized and incoherent depending on the number of particles and interaction strength. In particular, for filling factor less than one, the existing delocalization introduces imperfect localization in the primary lattice. In the strongly interacting limit, through fragmentation we establish that the many-body state is a fragmented superfluid. Introduced disorder tries to eliminate the superfluid layer and establish a insulating phase. The measure of order parameter establishes that disorder introduces a fragmented Mott. This highlights the fundamental role played by the secondary lattice to stabilize and create more coherent many-body state. We observe more interesting physics for filling factor grater than one, when the degree of localization of these extra particles is determined by the interaction strength. In the fermionization limit, Mott dimerization happens in the primary lattice- dimers with clear correlation hole is observed in the two-body correlation function. Disorder first pushes the dimer from one lattice to next lattice, without affecting the intra-pair correlation in the dimer, the background Mott correlation also remains unaffected. Stronger disorder is able to reduce Mott correlation. Our study explores some unconventional localization process in the strongly correlated systems.

From an experimental perspective, our predictions are
well within reach of current ultracold atomic experiments. State-of-the-art techniques allow for the realization of optical lattice of different filling factors, quasiperiodic optical lattices with precisely tunable disorder strengths. Our study suggests clear experimental signatures, such as density redistribution patterns and
correlations, which can be directly measured using quantum gas microscopy. This paves the way for future investigations for dipolar many-body systems.

\section*{Acknowledgments} 
BC acknowledges significant discussion of results with P. Molignini.
BC and AG thank Fundação de Amparo à Pesquisa do Estado de São Paulo (FAPESP), grant nr.~2023/06550-4. AG also thanks the funding from Conselho Nacional de Desenvolvimento Científico e Tecnológico (CNPq), grant nr.~306219/2022-0. 

\appendix

\section{ The Multiconfigurational Time dependent Hartree method for bosons}\label{multi}
The ansatz for the many-body wave function is the linear combination of time dependent permanents
\begin{equation}
\vert \Psi(t)\rangle = \sum_{\bar{n}}^{} C_{\bar{n}}(t)\vert \bar{n};t\rangle,
\label{many_body_wf}
\end{equation}
The vector $\vec{n} = (n_1,n_2, \dots ,n_M)$ represents the occupation of the orbitals, with the constraint that $n_1 + n_2 + \dots +n_M = N$, which ensures the preservation of the total number of particles. 
Distributing $N$ bosons over $M$ time dependent orbitals, the number of permanents become  $ \left(\begin{array}{c} N+M-1 \\ N \end{array}\right)$.
The permanents are constructed over $M$ time-dependent single-particle wave functions, called orbitals, as 
\begin{equation}
\vert \bar{n};t\rangle = \prod^M_{k=1}\left[ \frac{(\hat{b}_k^\dagger(t))^{n_k}}{\sqrt{n_k!}}\right] |0\rangle 
\label{many_body_wf_2}
\end{equation}

Where $|0\rangle$ is the vacuum state and $\hat{b}_k^\dagger(t)$ denotes the time-dependent operator that creates one boson in the $k$-th working orbital $\psi_k(x)$, \textit{i.e.}:
\begin{eqnarray}
	\hat{b}_k^\dagger(t)&=&\int \mathrm{d}x \: \psi^*_k(x;t)\hat{\Psi}^\dagger(x;t) \:  \\
	\hat{\Psi}^\dagger(x;t)&=&\sum_{k=1}^M \hat{b}^\dagger_k(t)\psi_k(x;t). \label{eq:def_psi}
\end{eqnarray}
The accuracy of the algorithm depends on the number of orbitals $M$.
$M=1$ corresponds to the mean field Gross-Pitaevskii equation.
If $M \rightarrow \infty$, the wave function becomes exact, with the set $ \vert n_1,n_2, \dots ,n_M \rangle$ spanning the complete Hilbert space for $N$ particles. However, due to computational limitations, the number of orbitals is restricted to a desired value to ensure proper convergence in the measured quantities. It is important to note that both the expansion coefficients $\left \{C_{\bar{n}}(t)\right\}$ and the working orbitals $\psi_i(x,t)$ that are used to construct the permanents $\vert \bar{n};t\rangle$ are fully variationally optimised and time-dependent quantities~\cite{TDVM81,variational1,variational3,variational4}. The time-dependent optimization of the orbitals and expansion coefficients ensure that the MCTDHB method can accurately describe the dynamics of strongly interacting bosons. MCTDHB has found extensive application in elucidating the statics and dynamics of various trap geometries and interaction strengths~\cite{rhombik_pra,rhombik_jpb,rhombik_pre,rhombik_scipost}. As the permanents are time-dependent, a given degree of accuracy is achieved with shorter time compared to a time-independent basis.
The time-adaptive many-body basis set employed in MCTDHB allows for the dynamic tracking of correlations that arise from inter-particle interactions.

\section{ Convergence in the initial density}
\begin{figure}
    \centering
         \includegraphics[width=0.6\textwidth, angle=270]{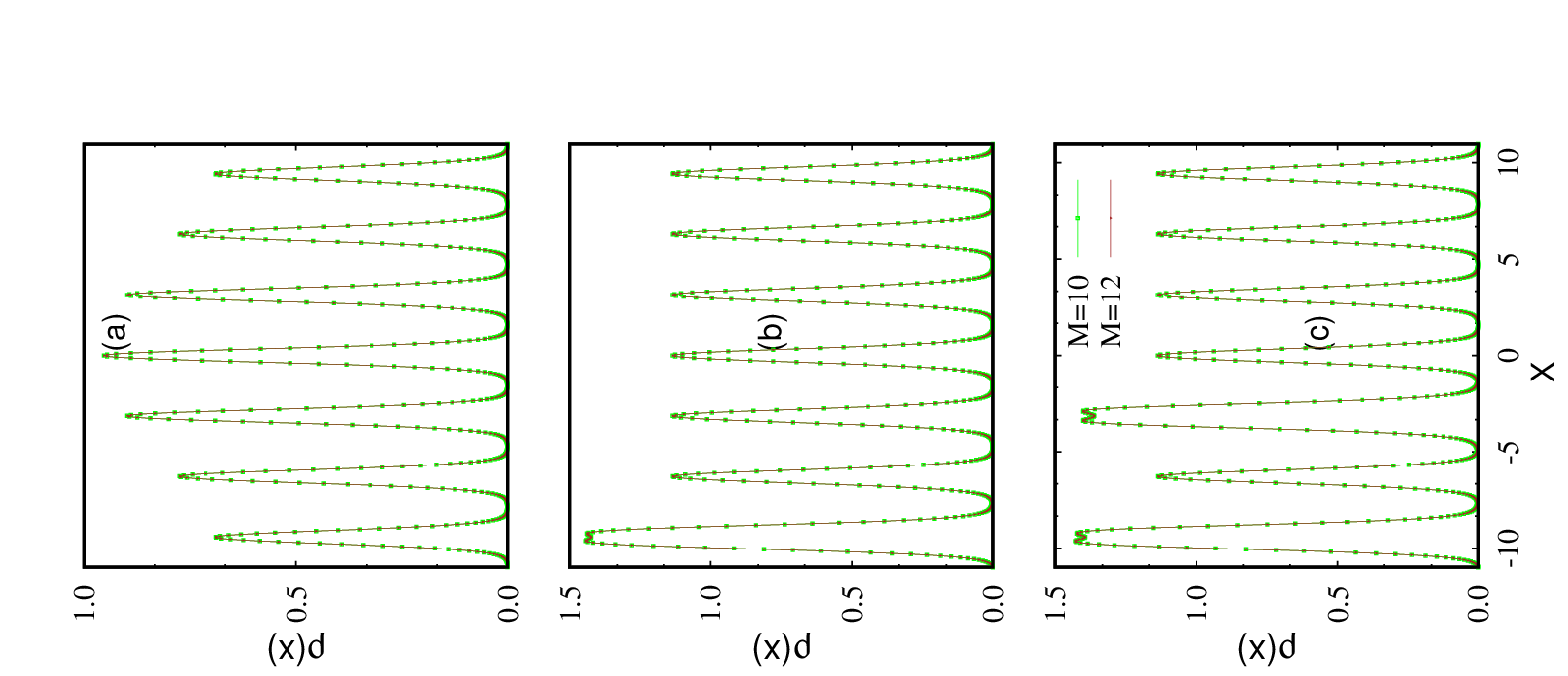}
    \caption{One-body density $\rho(x)$ for different orbitals ($M$) for seven lattice sites in the primary lattice and for different incommensurate filling.
    (a) corresponds to five bosons with $\lambda=5E_r$; (b) corresponds to eight bosons with $\lambda=10E_r$ and (c) corresponds to nine bosons with $\lambda=10E_r$
     }
    \label{fig9}
\end{figure}

The accuracy of MCTDHB ansatz strictly depends on the choice of the number of orbitals $M$ employed in the computation. In the limit of $M\rightarrow \infty$, the wave function becomes exact by construction and numerically it is achieved employing a finite number of orbitals for which the calculated observables exhibit convergence with respect to orbital numbers. 
In Fig.~\ref{fig9}, we plot the density profile for different incommensurate filling factor with increasing orbital numbers.  We specifically focus for the strongly interacting limit, convergence is established with $M=10$ and $M=12$ orbitals.   Fig.~\ref{fig9}(a) corresponds to five bosons in seven lattice sites with $\lambda=5 E_r$;  Fig.~\ref{fig9}(b) corresponds to eight bosons in seven lattice sites with $\lambda= 10 E_r$;  Fig.~\ref{fig9}(c) corresponds to nine bosons in seven lattice sites with $\lambda=10 E_r$. For all cases, density profiles with $M=10$ and $12$ are indistinguishable. Throughout our calculation we keep $M=12$ orbitals irrespective of filling factors.

\section{System parameters}

The quasi periodic lattice is the superposition of two optical lattices; a primary lattice of depth $V_p$ and wavelength $\lambda_p$ and a secondary lattice of depth $V_d$ and wavelength $\lambda_d$ parameterized as 
\begin{equation}
V(x) = V_p\sin^2(k_px) + V_d \sin^2(k_dx)
\end{equation}
where $k_i$ is the wave vector. We choose $\lambda_p \simeq 1032$~nm and $\lambda_d \simeq 862$~nm, which are compatible with real experimental realizations in ultracold atomic gases. These give vectors $k_p \simeq 6.088 \times 10^{6}$ m$^{-1}$ and $k_d \simeq 7.289 \times 10^6$ m$^{-1}$. We set hard-wall boundaries to restrict the optical lattice to the center-most seven minima.

\begin{table}[ht!]
\centering
\begin{tabular}{ || c | c || }
\hline \hline
Quantity & MCTDH-X units  \\
\hline \hline
unit of length &  $\bar{L} = \lambda_p/3 = 344$ nm \\
\hline
unit of energy & $\bar{E} = \frac{\hbar^2}{2 m \bar{L}^2} =E_r (\frac{3}{\pi})^2$\\
\hline
potential depth & $V=10.0 \bar{E} \approx \: 9.128 E_r$ \\
\hline
on-site repulsion & $\lambda = 0.5 \bar{E} \approx \: 0.456 E_r$ \\
\hline \hline
\end{tabular}
\caption{Units used in MCTDH-X simulations. $E_r=\frac{\hbar^2 k_p^2}{2m}$ is the recoil energy.}
\end{table}

\subsection{Lengths}
In MCTDH-X simulations, we choose to set the unit of length $\bar{L} \equiv \frac{\lambda_p}{3}= 344$ nm, which makes the minima of the primary lattice appear at integer values in dimensionless units, while the maxima are located at half integer values. $x=0$ is the center of the lattice which  can host an odd number of lattice sites $S$. In our numerical simulation, we consider $S=7$ sites and we run simulations with 512 grid points. \\

\subsection{Energies}

The unit of energy $\bar{E}$ is defined in terms of the recoil energy of the primary lattice, i.e. $E_r \equiv \frac{\hbar^2 k_p^2}{2m} \simeq  3.182 \times 10^{-26}$ J  with $m$ $\simeq$ 38.963 Da, the mass of $^{39}$K atoms.  Thus we define the unit of energy as $\bar{E} \equiv \frac{ \hbar^2} {2m L^2}$ = $ E_r (\frac{3}{\pi})^2$ = $ 2.904 \times 10^{-26}$ J. In typical experiments with quasi periodic optical lattice the depth of the primary lattice is varied in the around few tens of  recoil energies and the depth of the secondary lattice is varied around few recoil energy. In our simulations, we probe similar regimes: $V_p$= $10 E_r$ and $V_2$ $\in$ $\left[ 0 E_r, 3 E_r \right]$. The on-site interactions are kept fixed $\lambda=0.5 E_r$ for weakly interacting, $\lambda=5E_r$ for strongly interacting limit and $\lambda= 10E_r$  for fermionization limit, these values can be achieved in the ultracold quantum simulators.

\subsection{Time}
The unit of time is defined from the unit of length as $\bar{t}$  $\equiv \frac{2 m \bar{L}^2}{\hbar}$ = $\frac{2 m {\lambda_p}^2}{\hbar}$= $0.1307$ $10^{-4}$ s. = $ 13.07 \mu s$. 
\bibliography{ref}
\end{document}